**Title:** Coercion, Consent, and Participation in Citizen Science

**Authors:**

| | |
|---|---|
| Alison Reiheld | Pamela L. Gay |
| Dept. of Philosophy, Campus Box 1433 | Planetary Science Institute |
| Southern Illinois University Edwardsville | 1700 E. Fort Lowell, Suite 106 |
| Edwardsville, IL 62026 | Tucson, AZ 85719-2395 |
| ORCID: 0000-0001-5537-9060 | ORCID: 0000-0002-1797-3138 |
| areihel@siue.edu, (517)410-9638 | |

**Abstract:**
Throughout history, everyday people have contributed to science through a myriad of volunteer activities. This early participation required training and often involved mentorship from scientists or senior citizen scientists (or, as they were often called, gentleman scientists). During this learning process, participants learned how they and their data would be used both to advance science, and in some cases, advance the careers of professional collaborators. Modern, online citizen science, allows participation with just a few clicks, and people may participate without understanding what they are contributing to. Too often, they happily see what they are doing as the privilege of painting Tom Sawyer's fence without realizing they are actually being used as merely a means to a scientific end. This paper discusses the ethical dilemmas that plague modern citizen science, including: the issues of informed consent, such as not requiring logins; the issues of coercion inherent in mandatory classroom assignments requiring data submission; and the issues of using people merely as a means to an end that are inherent in technonationalism, and projects that do not provide utility to the users beyond the knowledge they helped. This work is tested within the context of astronomy citizen science.

**Acknowledgments:**
This work was in part funded through NASA cooperative agreement NNX17AD20A. Any opinions, findings, and conclusions or recommendations expressed are those of this project & do not necessarily reflect the views of the National Aeronautics and Space Administration (NASA). This paper would not have been possible without the space, internet, and unending coffee found at North Main Street Dinner, Edwardsville, Illinois.

**Introduction**

When is asking someone to perform a task in fact asking too much? This general ethical question about the limits of what people can rightly demand of each other is one that too often goes under considered. Little explored are the ethics of citizen science, and in particular citizen science in classrooms. Whether or not the conduct of citizen science is ethical depends on power relationships and what data researchers take from people, as well as how projects use data and acknowledge those who produced it. The need to review the ethics of citizen science projects is particularly relevant in the post-Cambridge Analytica (Isaak & Hanna 2018) and GDPR era (European Commission 2018), where examples of unethical data harvesting and use trigger public discussion.

  Citizen science as a phrase is relatively new, dating only to 1989 (Kerson 1989), but the concept is old: everyday people contribute specific data to a science project that otherwise would not be able to produce discoveries. In the earliest known example, farmers and citizens in revolutionary war era America contributed weather data in hopes better forecasts could be made — data that both had the potential to help win the war and improve their farming (NOAA 2006; NOAA 2019). Over time, people have collected data for projects needing large geographic, celestial, or time coverages. This includes measuring stellar brightnesses over time for the American Association of Variable Star Observers, both passing stars from observatory to observatory as the world turns and passing stars across generations as they evolve. Traditionally, participants were adults and concepts of informed consent, exploitation of volunteers, and utility of projects to participants were unapplied — largely because these concepts didn't exist as firm guiding principles of human participant research, even in medicine, until the mid-20th century (Code 1949; Department of Health 2014).

  While that first weather-related citizen science project is still active today, it and most everything else about citizen science has undergone a digital revolution that requires researchers to reconsider the ethics of how they present their projects and the potential for harm their volunteers accept in participating. Further, citizen science is often presented as a tool for increasing science literacy (Bonney et al. 2009), and in the U.S. as a means for decreasing the science costs, and thus as necessary to the nation's economic competitiveness (Gardner 2017). Such technonationalism can be used to justify all manner of exploitation, and has been used to do so in the Chinese context with fetal stem cell research and other bioscience research that is explicitly used to build Chinese technological capacity and advantage (Song 2017).

  In the academic field of ethics, these concerns are addressed through the concepts of utility, exploitation, consent, and coercion. Well-developed literature exists for the ethics of medical research, which is analogous to citizen science in this analysis: both involve human participants whose participation ranges from voluntary to coerced, and who can be exploited for the utility of national agendas. Like medical research participants, citizen scientists can be subjected to harm while receiving little or no benefit from their participation. One of the alleged participant benefits of participating in citizen science is enhanced education. Educational research on the efficacy of student research and authentic science experiments as tools for enhancing learning and self-efficacy, and effects on future learning and career choices, reveals uncertainty of benefit, especially in the context of disenfranchised groups (Roth & Lee 2004; Yaser & Baker 2003). Synthesizing these two bodies of literature puts citizen science into a new ethical context in which the concerns of exploitation, coercion, consent, and utility are very much alive.

This paper discusses several potential ethical pitfalls related to citizen science within the context of astronomy. This analysis looks at how project design affects volunteer ability to consent to data collection, and how communications of project goals can both mis-represent potential project utility, and hide potential exploitive characteristics. It is critical to emphasize reciprocity and genuine consent if exploitation is to be avoided, and to take into consideration the choices available to participants as well as the degree to which they understand the role they, and their data, will play in the career advancement of others.

**What (generically) is citizen science**

In this paper, citizen science is defined as science that is advanced through the engagement of volunteers from the general public; true citizen science has the ability to produce new knowledge only because the citizens were engaged. A project involving individuals in participatory science, but that does not have the potential to advance research, is authentic engagement or participatory science, but is not citizen science. Table 1: Citizen Science versus Participatory Science provides parallel examples of these engagements.

By its nature, citizen science engages citizens as knowledge producers, not simply as knowers. The term "knower", from the philosophical discipline of epistemology, refers to the person who acquires knowledge and has it. Knowledge producers, by contrast, generate new knowledge which they and others may acquire, thus becoming knowers. Citizen scientists are both knowers and knowledge producers. Participatory science involves only knowers. This distinction has implications for giving credit where it is due and for the use of data.

The ways in which citizen scientists can participate are highly varied, but generally group into three broad (and combinable) categories: donating facilities (a backyard weather station, unused computer cycles), donating observing time (monitoring variable stars, tracking asteroids), and donating analysis (mapping features in images, identifying patterns in graphed data). Sister efforts, such as software development, project translation, and community moderation, aren't citizen science but enable it.

**What is utility**

The research teams behind citizen science projects often argue that their projects have utility to humanity at the individual and the society level. On the community scale, projects may document light pollution that affects the individuals collecting the data. On the global scale, projects may help to determine the likelihood of asteroid impacts or map out worlds humans may someday walk across. Additionally, in terms of individual utility, science teams promote their projects as meeting a variety of science education and science literacy goals.

The concept of utility has many variations. In its most basic form, as John Stuart Mill (2016) said, an action takes its value from the purpose it serves. Utility is fundamentally a consequentialist concept: whether a course of action is good will depend on its consequences — its utility — for achieving some purpose. If one's purpose is technological ascendency of a nation, utility may be very different from that which serves the purpose of achieving human dignity or self-efficacy. Whenever someone asks about the utility of an action, they are judging consequences and should in turn ask, with respect to what purpose? And is that purpose *Good*? In citizen science, it is imperative to ask, with what purpose do researchers engage volunteers? And are they taking into consideration consequences for volunteers?

For Mill, perhaps the most famous proponent of the ethical theory of utilitarianism, the Good served by right actions is to maximize happiness and minimize suffering. Actions are right in proportion as they increase happiness and decrease suffering, and wrong as they tend to the reverse. This is known as the Greatest Happiness Principle. Happiness here is not some basic notion of mere pleasure. Rather, for Mill, pleasures come in both intellectual and animal forms. The intellectual pleasures are "higher" pleasures, such as discovery, art, invention, and debate, while lower pleasures include basking in the sun, reproducing, eating for survival, and so on. Higher pleasures are to be valued over lower pleasures, or as Mill said, better to be Socrates dissatisfied than a pig satisfied. Scientific investigation and discovery, and that joyous feeling of the "Eureka!" moment, would certainly be a higher pleasure and preferred over the lower ones, thus worth some physical suffering and delayed gratification. With volunteers, projects need to consider whether they serve these higher pleasures or cause harm. *Policies which do not adequately consider the consequences for citizen scientists in calculations of utility are simply unacceptable.*

*Citizen Science: A means to an end*

At the most fundamental level, citizen science exists to benefit the scientist and is a means for completing tasks that could not be completed without the aid of volunteers. So why do volunteers volunteer? In studying the motivations of citizen scientists, it quickly becomes clear that people are interested in helping, and are hopeful they will discover something (Raddick et al. 2010; Bakerman 2018). Acknowledging this is one way to account for citizen scientists in calculations of utility. But is it enough? The combination of the scientists having work to be done, and the volunteers being motivated to do it, can create a Tom Sawyer situation, where scientists talk people into doing their busy work for them without fully articulating how they might (or might not) benefit. Tom's actions are unethical because he dupes his friends into thinking painting is a privilege without saying that he will benefit from their efforts and suffer if they don't help. Had Tom asked his friends to "do him a solid" and help with the painting so he wouldn't be punished, then their satisfaction in helping would have been an honest reflection of what they had done. Instead, the duplicity of Tom's actions tricks them into a false sense of pleasure. The happiness generated by it will dissipate if the truth is revealed. Along these lines, if a researcher asks citizen scientists to complete a task, like mapping galaxies, that the researcher could do but doesn't want to, and convinces the public to do it because it is a privilege to participate in scientific discovery, will the participants feel tricked when they see the researcher get publications and prestige on the backs of their effort while they have received only the "privilege" of contributing their labor?

Not only is it critical to be truthful in order for the pleasures of citizen science to be genuine and sustainable, but it may be incumbent on researchers to provide additional utilities to citizen scientists. The most typical form that these take in citizen science are increasing the scientific literacy of the individual and increasing STEM engagement by society, and projects that facilitate learning also see that as a motivation to their audience (Gugliucci et al. 2013). One ethical path forward in citizen science is to explain exactly why citizen efforts are needed, who will benefit from their efforts in what ways, and to take care to support higher pleasures such as learning. This kind of purposeful care can avoid the Tom Sawyer-ism that is starting to become embedded in U.S. legislation and science funding.

In an industrial society, citizens are often seen as a resource that can be used. This was articulated as part of legislation in the 2017 *American Innovation and Competitiveness Act* (Gardner & Rept), where it states in §402.b.2 "crowdsourcing and citizen science projects have a number of additional unique benefits, including accelerating scientific research, increasing cost effectiveness to maximize the return on taxpayer dollars, addressing societal needs, providing hands-on learning in STEM, and connecting members of the public directly to Federal science agency missions and to each other," and goes on to grant federal agencies authority to use citizen science to advance science. This law requires that citizen science practitioners comply with part 46 of Title 28, Code of Federal Regulations (Dept. of Justice 2011), and conduct their experiments under the guidance of Internal Review Boards, which will make sure that no direct harm occurs to participants. Taken in its entirety, this law looks to address societal needs and provide learning, which are, as discussed above, a higher utility. Unfortunately, citizen science projects aren't required to fulfill all aspects of this law, and it does not prevent a very particular kind of wrong which is always possible when humans are seen as resources: that they may be treated merely as a means to an end.

Here, this investigation moves from utilitarian considerations of harm to considerations of hard limits on the ways in which one may permissibly use humans as resources. In research ethics concerning experiments with human participants, the *sine qua non* is full and informed consent, that the person willingly *gives themselves up for use* as a resource. It is this principle, not only utilitarian concerns of mitigating harm, that underpins the earliest restrictions on human research as codified in the Nuremberg Code, a direct ancestor of the Belmont Report, itself the founding document of American federal regulations known as the Common Rule (1979).

To use someone as a resource is not prohibited under the Common Rule, nor, as the reader shall see shortly, is it prohibited under Kantian ethics. Students use teachers as resources every time they learn. Authors use journal editors every time they seek to publish an article that will secure their employment. And medical researchers use even the most willing of human participants as a means for discovering treatments that may only benefit future patients. To paraphrase a famous ethical argument (Thomson 1971), the minimally decent standard one must never fall below is that of never using people *merely* as resources. Rather, one must always remember that it is a person who is the resource and that the use of resources who are persons is very different from the use of resources that are things. For researchers, this means always keeping the value of the participant foremost in their minds, above even the value of the research goal.

The alert reader will note a common feature of this person-centered analysis and the utilitarian analysis: citizen scientists must be centered, with their needs and interests considered and attended to. Though their labor is what researchers seek, researchers must always be mindful that labor is produced by laborers.

*The utility of citizen science as a means to scientific literacy*

One common way to address the needs of project participants is to educate them. There is unlimited potential for citizen science to create an apprenticeship model in which volunteers learn the methods and knowledge of the scientists running a program. This is commonly seen in observational programs, such as those associated with the American Association of Variable Star Observers and the 2017 Solar Eclipse related Citizen CATE program (Penn 2017). These programs have built into them training and the chance to learn the details of the science. They often include forums where all members of the project, from scientists to volunteers, can collaborate together. Other projects, like CosmoQuest, embed educational content, community engagement, and collaboration within their fully-online citizen science facilities.

These models work to maximize intellectual utility to participants while allowing discovery to potentially take place. While people are not required to learn to participate, these projects proactively invite participants to learn, making it clear they want to do more than use people as a mere means to an end.

Other projects allow space for learning without explicitly making it part of the project, and research indicates that people engaged in citizen science may self-select to learn about the science they are engaged in (Masters et al. 2016). These kinds of projects, such as Zooniverse projects or the Great Backyard Bird Count and eBird, engage people in pattern matching and related tasks (counting birds, mapping features, tagging structures). While it is not necessary to understand the science of these projects to participate, participation may inspire people to want to understand the science. The question than becomes, is that enough? Is asking people to complete a task that may have the utility of inspiring learning, but that isn't designed to inspire learning, sufficient? We would argue that while this may be sufficient for the standard of not treating people merely as a means to an end, it is not sufficient to satisfy utilitarianism's Greatest Happiness Principle. Of the available courses of action, that action which maximizes happiness and minimizes suffering for all concerned is the one which provides participants with a wide variety of utilities, including projects designed to inspire learning, not merely to make it possible.

Ultimately, to be sure we are fulfilling obligations to our citizen scientists, it is imperative that we talk to people and understand their perceptions and their needs. This goes beyond asking what motivates people to do citizen science. We must also ask, what is the perceived utility of this work to the participant, and what do they require to not feel like a mere means to an end?

*The utility of citizen science as a means to benefit society*

Beyond considerations of personal utility are those of community and societal utility so as to maximize happiness and minimize suffering for all affected. Increasing STEM engagement falls into this category. In the United States of America, Congress and the White House have determined that increasing STEM engagement is paramount for national competitiveness (Gardner & Rept 2017; Committee on STEM Education 2018). This is a form of technonationalism.

The ideal behind technonationalism is to create a population whose people are engaged in STEM-focused education and employment. The subtext of this ideal is that people who aren't STEM-focused are less advantageous to society. Within the context of citizen science, the argument becomes, it is in the national interest for people to engage in citizen science to help advance science at lowered costs. It is assumed these people will learn science (the research is less clear).

This technonationalist argument ignores that there are many different ways to advance STEM, and in some cases, necessary engagement includes being a professional glass blower in a chemistry lab, a plumber in an aquarium, or the Human Resources Specialist at a research center. These people are pursuing necessary crafts that support STEM fields, but may not be considered to be part of the STEM workforce. Taken further, it is necessary to have non-STEM professionals to advance STEM.

If advancement of STEM doesn't require everyone to be STEM-focused, and if participation in citizen science doesn't necessarily promote STEM learning or STEM careers, then the only utility to society is to lower the costs of doing science by using people merely as a means to an end. Further research is needed to learn if society sees this as sufficient benefit unto itself. But on a utilitarian analysis, perceived benefit is not the same as actual benefit. It is the actual sum total of happiness and suffering that utilitarianism would consider definitive. This ambiguity in there being sufficient utility highlights the need to make it clear to participants how they / their data are being used, to gain consent for that usage, and to strive to find ways to provide higher utility to volunteers such as through educational engagement.

**Consent in Citizen Science: participation and data-sharing**

As previously discussed, informed consent and refusal is the *sine qua non* of research ethics. In citizen science, we must consider two very different contexts of consent: whether a volunteer is fully able to freely choose to participate or not participate, and whether they are able to provide informed consent to have their data used. Tangentially, we also run into issues of funders and oversight bodies requesting or judging projects based on their ability to collect demographic data that demonstrates projects meet the stakeholders' goals, even though these may have no relation to the projects' science goals.

      The concept of consent has a long history in ethics as an important corollary of voluntariness. In ancient Greece, Aristotle discussed the importance of voluntariness for assigning blame or praise for actions. He said that if someone were coerced into performing an unethical action by threat of force, we would not consider it fully voluntary and therefore would assign less blame. In the modern Western context, German philosopher Immanuel Kant prized autonomy and rational choice as a feature of what it means to be a rational being. Because of this, rational beings have an inherent value completely independent of any use to which they may be put. To reduce a person to "merely a means to end" rather than treating them as an end in themselves — a creature with value that *cannot simply be used as a tool* — is, for Kantians, an utter violation of morality. The reader will recall our previous discussion of limits on the use of participants as resources. It is Kantian ethics that often gives the moral force to these limits. While people can of course use each other as a means to an end — as students use teachers, as authors use editors, as all medical research uses participants — humans must not use each other *merely* as a means to an end, as when research is conducted without the knowledge and consent of participants. This consensus view was developed in Europe and North America in response to Nazi medical experiments (Lifton 1986; Utley 1992), codified in the Nuremburg Code, and tested in North America by the Tuskegee syphilis study (Brandt 1978; Pellegrino 1997) which began prior to the Nuremberg Code but which continued even after it was crafted and disseminated. We also see this principle in the Belmont Report's famous principle of autonomy: that persons should be self-governing and be allowed to not only consent to but also to refuse to participate in research. *Consent without the real possibility of refusal is not genuine consent.*

      As Aristotle long ago noted, voluntary actions, such as consent, can be undermined by coercion. Aristotle's example involved the threat of use of force, but medical research considers undue compensation or threats of loss of access to medical care or longer prison terms to all be coercive in the extreme. In 2016, a judge in the state of Tennessee offered prisoners early release in exchange for "consenting" to sterilization or long-acting contraceptives. Note that while prisoners technically were given a choice, the external incentives would powerfully drive them to choose sterilization or long-acting contraceptives. Psychologists refer to the context in which choices are made as "choice architecture", and medical ethicists have begun paying close attention to the ways in which deliberate structuring of choice architecture can be coercive (Blumenthal-Barby and Burroughs 2012). *Choice architecture that leaves only one reasonable choice cannot yield genuine consent.* Rather, it is coercion.

What bearing does this have on citizen science? By their nature, citizen science projects expect individuals to contribute information necessary for science and sometimes for external stakeholders (e.g. funders). This information is then used in ways that may or may not be fully understood by volunteers (an issue of informed consent), and some 'volunteers' may participate as part of classroom, association, or required volunteer activities (an issue of coercion). For the scientists, their primary goal is to get information so they can complete research — a context that risks using humans as merely a means to an end, reducing them to tools. For external stakeholders, goals may be more bureaucratic in nature such as demonstrating that a key audience dominates a volunteer pool or achieving technological ascendancy as a nation; humans may be reduced to group membership (e.g. specific geographic populations) and how they check boxes for representativeness of a study population. This latter goal is in service not only of higher quality studies that can better be generalized to the population as a whole, but also in the service of fairness and inclusion of folks who are too often left out of both knowledge — i.e. Lyerly, Little, and Faden (2008) on the exclusion of pregnant women from research — and knowledge production.

For this reason, citizen scientists are asked to contribute not only scientific, but also demographic information. Researchers conducting citizen science not only produce scientific knowledge, but also seek to produce databases of information about citizen scientists and their behavior. This can be — and often is — used to study the behavior of volunteers and efficacy of citizen science.

To do so, researchers step back from claims about the natural world to consider claims about the people producing knowledge about the natural world. This, at least as much as the production of scientific knowledge, also risks using humans as merely a means to an end, reducing them to not even tools, but to data points. And it does so on the basis of personal information that citizen scientists may or may not be aware could be used in this way.

*Informed Consent*

In citizen science, consent is required for two different sets of data. First, citizen scientists must consent to have the science data they submit used for science research. Second, they must consent to have any potentially personally identifying information — age, and other demographics — used to produce population research about citizen scientists. There is a separate issue of allowing them to submit a publishable name that can be used in press releases, scholarly publications, and other venues to give them credit.

The ethical issue here is not with the goal of understanding who is involved in knowledge production. That may be necessary to satisfy moral principles such as justice, fairness, and inclusion. Rather, the issue is that in citizen science, information-gathering has twin objectives: to produce scientific knowledge, and to produce knowledge about knowledge producers. At best, our informed consent processes focus on making citizen scientists aware of the former. Too often, they leave out the latter. The reduction of citizen scientists to mere means to an end — however noble that end — occurs whenever the consent process does not produce genuine autonomous authorization. For Ruth Faden and Tom Beauchamp (1986), autonomous authorization of a medical intervention does not involve merely checking off the legal boxes required for consent, but achieving all of the following criteria:

1. The consenting person has *substantial understanding* of the information they need in order to make their decision

2. The consenting person acts *in the absence of coercion*
3. The consenting person *intentionally…*
4. *…authorizes/consents* to allow a professional to do the intervention

For our purposes, we will focus on the criteria of *substantial understanding* and of *the absence of coercion*.

If either of the twin objectives of citizens science information-gathering is not clear to the participant, genuine autonomous authorization has not been achieved. If the choice architecture is coercive because the participant cannot refuse without penalty, genuine autonomous authorization has not been achieved. As we shall see in our discussion of the ways that citizen science is used in the classroom, this is a genuine concern.

Research into public understanding of science has long shown that most people do not fully understand either the methods of science nor the results of science. In particular, they often do not understand how research results are disseminated, how credit for knowledge is assigned, and how that credit can benefit the careers and even salaries of professional scientists. We are concerned that citizen scientists who are science-savvy about the results of science may nonetheless not fully grasp what they are consenting to about the use of their work when they contribute to knowledge production. What's more, citizen scientists producing scientific data for projects that are internet-based may have widely varying language skills due to language of origin, age, and educational level, as well as communication differences based in perceptual and social differences, along with other factors. The quality of informed consent, and thus whether it is ethically adequate, depends on citizen science understanding of explanations of process and method. This understanding in turn depends on both citizen scientist background knowledge and on ability to comprehend researchers' attempts to fill in that background knowledge.

Collection of scientific data has the potential to directly benefit citizen science by providing scientists (and the world) with new scientific catalogues and scientific understandings that they find interesting. Observations of users participating in usability testing of the CosmoQuest website showed that all users understand their data would be used to do something, even if they didn't correctly understand what that something was. Motivations research finds volunteers' primary motivations for participating are largely to help/contribute to science or to make discoveries. Anecdotally, when the Uwingu[A] project launched, the CosmoQuest project received a noticeable amount of mail from people concerned the craters they mapped might have their naming rights sold, and were willing to sue to prevent this from happening. These observations indicate that people want to contribute even if they don't fully understand how they are contributing, and that a vocal few potentially reflects a larger population of volunteers who take ownership of what they do. We believe it is necessary to clarify how the volunteers will and will not receive attribution for their efforts from the beginning. For instance, they will be part of a long list of names on a website that is consistently linked to, but press coverage will cite the project scientists and most likely will not cite any volunteers by name. This is consistent with publication agreements found in many large collaborations, and also set volunteers up to not feel wronged when their name isn't specifically listed on every project publication.

To do this correctly, a link to terms of service, or even a brief block of clean and concise text may not be sufficient; there are people who will not read text, and also volunteers who may have followed a link to the site but not actually speak English. To allow for ethically adequate informed consent, medical settings sometimes use wordless videos. This is especially useful for showing what is involved in various medical procedures. However, this model does not lend itself to our context. We thus follow another model that is starting to appear in the medical community by using cartoons without words to represent ideas and questions to non-verbal and non-English speaking patients. (See Figure: Informed Consent Graphic).

Collection of demographic information is in many ways harder to explain and justify to participants. Demographic information is generally collected for two reasons: to satisfy the needs of funders who are concerned about who the projects they fund interacting with, and by human-subject researchers who are studying who does citizen science. Research into how to effectively communicate these secondary lines of research is needed.

To address these hurdles, we propose an informed consent process during registration that is:
1) Roughly but adequately explanatory in purely visual iconography and flowcharts with almost no text
2) With additional English-language, plain-language text explanation available
3) Aimed at explaining not only what we specifically do with information but also at filling in background knowledge about how information may be used by scientists
4) Addresses both scientific information produced by citizen scientists used in science research, and information about citizen scientists used in population research about citizen scientists

Since one of our concerns with citizen science is inclusiveness, iconographic representations of users and researchers should disrupt the stereotype that science is done by white men. Icons should be designed to expand the variety of people in professional science, showing science as a viable career option or lifelong involvement in citizen science as a real possibility for diverse audiences. During the image design phase, for instance, we considered representing scientists as wearing top hats in order to disambiguate them from citizen scientists. However, this kind of hat was historically worn almost exclusively by upper-class white men. We discarded this in favor of a cliche but obvious lab coat.

Of course, no amount of understanding can produce ethically adequate informed consent if that consent is not truly free but is rather subject to undue coercion that compromises the very notion of voluntariness.

*Coercion and Data Use without Consent*

In June of 2014, a group of researchers published a paper on emotional contagion based on research using Facebook without the knowledge of the user base that participated (Kramer, Guillory, and Hancock 2014). The experiment was simple: they showed people preferentially happy or sad posts and then looked to see how their emotions were impacted as revealed by reader responses. In the time since news of this experiment broke, the concept of informed consent has gained traction in popular discussions as both researchers (e.g. Flick 2015; Shaw 2016) and bloggers have filled the internet with papers and think pieces on when it is and is not appropriate to use people without consent in experiments implemented through social media. In combination with Facebook's mistake of giving away data to Cambridge Analytica (Granville 2018), this social media giant has brought two major concerns into the public dialogue - when is it ok to use people's data without their consent, and when is it ok to do research on people without their consent? For many, the answer is never, especially *but not only* when that use may do harm.

In citizen science, it is all too easy for researchers to find themselves on the same path of ethical missteps as Facebook. The data submitted to projects can be used by scientists to do cutting edge research that can bring the scientists promotion and reward. The patterns of when and how data is submitted makes it possible to study the behavior of volunteers in response to variables like site design, social media, and more. If a project explains to its participants how it uses collected data, and has a population of willing volunteers, the project should be able to avoid ethical problems. Issues arise, however, when people don't provide consent, are coerced into participation, or don't fully understand how their data can be used.

Understanding isn't required for projects to succeed. After all, it is possible for projects and scientists to benefit from the contributions of people who don't grasp how they are contributing. In 2016, the CosmoQuest project worked with a third-party website usability evaluator. Per the industry practices of the time, the project was instructed to reduce and remove the majority of website text because it is known that people are less likely to click through links when faced with a lot of words. During design testing in both 2017 and 2018, users testing projects to map the Moon, Mars, or Mercury were asked what they thought the purpose of the site was. Due to the lack of words, none of them knew, even though the tutorial trained them to accurately complete tasks. This created a choice: CosmoQuest could follow the best practices in design, and remove text to maximize the number of users who click through to map, or it could do what was ethically correct and keep the text and the uglier design to improve the likelihood people would understand what they were doing. Unfortunately, funding-related mandates required CosmoQuest to use the low-text design; the project was required to follow all recommendations from the consultant. As a result, the majority of the project's existing users left, and new users, while clicking through to try projects more often, didn't stay. The combined result was a significant reduction in project participation. This was reversed in 2019, with the launch of Bennu Mappers and a return to a more text-heavy design that delineated how data would be used and credit would be given. This correlated with a greater duration of project participation.

Ethically, projects are better positioned by making sure their volunteers understand how both they and team scientists can benefit. Research shows volunteers are motivated by the potential of making a discovery. If a participant doesn't have substantial understanding of how they will or won't receive credit for their contributions prior to undertaking the task, this motivation can lead to hurt feelings or worse if a volunteer learns they contributed to a major discovery and there is a disconnect between what they expect to happen (in some cases, seeing their name listed as a co-author), and what actually happens (in some cases, no credit is given by name/username at all). Not only is betrayal of trust harmful, but it violates perceived promises and is arguably exploitative. This kind of wronging of the citizens scientist is to be avoided, and can be avoided simply by spelling out how project credit is given to contributors. Volunteers may also find themselves distressed if they learn that their contributions personally benefit team scientists, by helping them gain career promotion, awards, or other things of value that aren't shared with the volunteers. Again, harm can be avoided by providing volunteers with upfront knowledge of how others may benefit from their efforts. This information gives participants the freedom to decide not to participate if they don't agree with how they and others will benefit (or not).

It is not enough, however, to merely post Terms of Service and Privacy Policies. One extra step is needed — volunteers must give consent, acknowledging that they agree to those terms and policies. While every possible effort should be made to motivate people to view these documents — and again, we are working to develop visual forms to allow understanding without requiring significant reading — we must point out that not everyone will view these documents. This is a widespread problem with all contexts of consent, from consent to participate in research to consent to receive medical treatment and tests. We know that many people won't take the time to understand what they are agreeing to. However, many will. We also know that it will be difficult to ascertain whether people actually do understand even if they have taken the time to try. Further research into ascertaining genuinely informed consent is needed. But now, with what we know, it is still important to require people to agree and make every effort to make substantial understanding possible. Regardless of what participants choose, it is wrong to simply take their data without seeking genuinely informed consent.

There are many projects, most notably Zooniverse-related projects, that don't seem to require login in order for people to participate. This is an unethical use of data because people may not know what they are contributing and have not consented to have their data used. Unfortunately, requiring login — a way to acquire consent before proceeding — will reduce the number of people participating in a project. While a minor barrier to participation, it is a barrier nonetheless. This is a necessary cost of taking a more ethical path; good data isn't good data if it isn't Good data.

Moving beyond volunteers contributing without giving explicit consent are issues related to people — most often students — being required to participate in citizen science projects. Recall that consent can only take place in the absence of coercion. A choice architecture that does not allow for refusal to participate, and requires consent, is one that is profoundly coercive.

Consider the common case of children being required to participate as part of mandatory classroom assignments. If given no other way to earn those credits, these students are being coerced: failure to participate has real and unavoidable negative consequences for their grades. At the same time, researchers — and the teachers whose reputation rests on student involvement in active learning — have the potential to benefit from these forced contributions.

As previously discussed, the impact can be a betrayal of trust experienced by citizen scientists. Teenagers and budding science enthusiasts can experience long-term damage to their interest in, and trust in, the processes of science when they experience this kind of betrayal during their formative years. If one of their first experiences with professional science is to be reduced to merely a means to an end, a cog in a machine, a tool to others' advancement, trapped in a set of choices that leaves them only the option of damaging their academic standing or disingenuously consenting to the project, we should be unsurprised when the result is distrust in science. We should also be unsurprised to see unchanged the very same pipeline issues we currently see with pursuit of STEM careers. On both Kantian and utilitarian grounds, such forced participation is wrong.

Furthermore, requiring students to do citizen science as part of their mandatory activities has not been shown to benefit student learning more than other activities (Gray et al. 2012), such as participatory science where students use well-scaffolded, open-ended curricula to perform experiments with known potential outcomes. It is even unclear from the research whether mandatory participation in research projects, such as through required Science Fair, consistently improves student performance and interest in science and science careers (Roth & Lee 2003; Yaser & Baker 2003). Without strong evidence that the benefits outweigh the harms of forced participation, there is even less reason on utilitarian grounds to require it. Put simply, students should not be required to do citizen science as part of mandatory coursework.

Teachers might be tempted to think, as we initially did, that a solution to this problem is to allow students to choose between citizen science and participatory science versions of the same project. However, this exposes students to both internal and external shaming for their choice. Instead, we recommend that the concepts of citizen science not be introduced beyond asking students to complete an anonymous tutorial while being introduced to the concept of citizen science. Teachers may in good conscience direct students to citizen science projects as optional assignments within a full suite of options. These concerns do not apply to elective courses or after school programs where students can also pick from options which are all genuinely optional. In such cases, schools and organizations such as Girl Scouts should use citizen science participation as one way to engage students in STEM opportunities.

Coercion can also be an issue for people who have mandatory volunteer hours and find that the only efforts that fit their limitations are citizen science related, whether online or in the field. This is both incentivized participation, and coerced because of the costs inherent in not participating. The CosmoQuest project has signed off on volunteer forms for people receiving conditional disability support who we learned through discussion did in fact have other options. Had these persons not had other options, we would have faced a moral dilemma. On the one hand, they clearly would not have genuine choice. This is morally bad, and accepting their consent in such a scenario is morally questionable. On the other hand, persons with social-emotional and mobility constraints often find it difficult to travel to a volunteer site and/or work in person with others. Online citizen science is an often asynchronous form of volunteering one's time that doesn't require leaving a space defined by physical or emotional constraints related to one's disability. It is thus morally good to welcome persons into citizen science who can garner very real material benefit from participation. It is unclear how often this issue may arise, but it is important to consider a variety of choice architectures in any analysis of informed consent and refusal.

By considering a variety of choice architectures, we are able to see otherwise invisible impacts on citizen scientists who have a variety of interests and limitations. One feature learned from the CosmoQuest online community through discussion boards is that many persons with social-emotional disabilities find this community provides them with social interactions that they find safe and supportive. What position are these folks placed in if they must consent to data sharing in order to participate in community? If refusing to share data means one cannot participate in the community, then perhaps refusal is not a genuine possibility.

Again, we face a dilemma between maximally autonomous consent and providing opportunities that may not exist elsewhere. This points to the need to allow community participation to take many forms, including those that don't require citizen science, and reinforces the need for these projects to provide means of higher utility such as educational opportunities. In the long run, this dilemma is best resolved by a society which provides many different ways of being in community; a truism to applied ethics is to look not only to changing individual behaviors but also systemic structures that set people up to fail. In the short run, however, participants with social-emotional disabilities need to be allowed the agency to trade off greater autonomy for increased community.

Beyond considering some of the reasons to engage in robust informed consent processes for both how results and personal demographic data will be used, further consideration should go into harvesting demographic data, and into acknowledging that people may wish to preserve their privacy. Beyond a general right to "informational privacy" which may exist (Whitbeck 2011), the ability to harvest large amounts of data about persons and their behavior raises the prospect of a new kind of privacy that protects people from being treated as mere targets for marketing or other manipulations. Philosopher and legal theorist Anita Allen has argued for a concept of "dispositional privacy." This allows one to keep private one's dispositions or states of mind, which can be inferred from aggregates of purchasing and other on-line behaviors as well as information about age, sex, and income. Classic examples of violation of dispositional privacy include the sale of databases containing datapoints people might not normally care about keeping private but which, when aggregated, allow the construction of a picture about them which they may not wish to be shared or sold.

It is especially important that projects make clear the ways that information gathered about participants may be aggregated and used. Ideally, projects will neither exploit nor facilitate the exploitation of participants. Consent is only one way to reduce the likelihood of exploitation, and the sale or manipulation of aggregate demographic data is only one possible form of exploitation. We now turn to a richer consideration of the ways that exploitation as a form of injustice can occur in citizen science.

**What is exploitation**

Exploitation is a common concept in everyday ethics. Worry arises when employers demand work beyond contractual requirements without additional compensation. Worry arises when the mineral resources of developing nations are extracted for the benefit of colonial powers. And worry arises when human research subjects are used in knowledge production without adequate informed consent. Researchers generally understand the plain language meaning of exploitation: to exploit someone is to use them without benefit or permission for the gain of others. But it will benefit the community to get even clearer on exploitation examining how it rears its head in citizen science projects.

Philosopher Iris Marion Young developed a framework characterizing what she called the "five faces of oppression": violence, marginalization, cultural imperialism, powerlessness, and exploitation. The face which citizen science must be most on guard against, is exploitation.

For Young, oppression is a structural form of injustice that affects individuals because of their membership in social groups, which often exist in asymmetric power relations. Groups, Young says, exist only in relation to other groups. The group identity of "citizen scientists" is necessarily defined by contrast with the group identity of "professional scientists." Groups do not always oppress other groups, of course. However, when oppression occurs, it may manifest in one or more of Young's five faces. As Young puts it, exploitation is characterized by "a steady process of the transfer of the results of the labor of one social group to benefit another." More than this, exploitation involves established "social rules about what work is, who does what for whom, how work is compensated, and the social process by which the results of work are appropriated", rules which enact relations of power and inequality (Young 1990). How, then, does this apply to citizen science?

Citizen science projects have the potential to exploit community members by using them as a means to produce science without providing benefit to the participants. While it can be argued that doing the science provides a benefit - that happiness in the form of higher pleasures is found in the journey - this argument is made by the people requesting data from volunteers and a more critical consideration is needed.

In theoretical ethics, both utilitarianism and Kantian ethics find exploitation morally problematic. Under utilitarianism, exploitation might be justified if it were the *only way* to satisfy the Greatest Happiness Principle. However, this is rarely the case. Nearly always, happiness is maximized and suffering is minimized by avoiding exploitation. This is because of the great suffering exploitation causes to some people in order to produce small happiness for others. Kantian ethics will not tolerate exploitation under any circumstances, for it violates the ethical requirement that we never use others merely as a means to an end. To those who would propose to exploit a small number of people for the benefit of the many, Kant would say "what part of 'never' did you not understand?" *Using others merely as a means to an end, and without consideration of the consequences for them, are definitive features of exploitation.* We are unsurprised when we find that coercion and exploitation go hand in hand, that genuine informed consent is nowhere to be found in cases of exploitation, and that simplistic considerations of utility are all too likely to result in exploitation.

Asking volunteers to contribute to a project without any potential personal benefit beyond the knowledge that they contributed risks using volunteers merely as a means to an end. In such cases volunteers are being exploited for their ability to contribute data. This can be avoided by giving credit to participants and by working to provide volunteers with benefits of higher utility, such as educational opportunities. Unfortunately, other forms of exploitation are not resolved by such means.

*Funding Exploitation*

In the highly competitive, modern funding environment, the question for any researcher is often "How do I force my research to fit this funding call?" rather than "What funding call is most appropriate for my research?" For citizen science, this takes on a novel form, as researchers shoehorn science projects into citizen science to take advantage of funding ear-marked for public engagement. There is also a compulsion to use citizen science to replace efforts that previously were the domain of students and junior researchers, as funding for junior positions evaporates.

This is driven by the America Competes Act, which demands that citizen science be used to reduce the cost of science, always attempting to do more with less. This approach clearly benefits researchers seeking to match more funding calls, and those pursuing technonationalist goals on the cheap. This shift in how projects are completed is of less clear utility to the volunteers who replace the students and junior researchers.

Historically, students and junior researchers accepted low pay for their efforts in exchange for advancement to the next stage in their career. These positions served as starting points for those who would be future leaders in their field; providing economic, intellectual, and career benefit. Unpaid volunteers do not receive these exchange benefits. Under America Competes, scientists are called on to use volunteers as merely a means to an end. Like so many funding structures, the result is a choice architecture that rewards researchers for treating citizen scientists as tools. Ethically, this is unacceptable, and projects should always ask, "How can engagement benefit our volunteers rather than simply benefitting us?"

While beyond the scope of this paper, researchers should also ask whether they are doing harm to the profession by replacing early career positions with volunteers not seeking to be professional researchers. Is it possible to expand public involvement in science without harming the STEM pipeline by removing one of the segments that has so long been key?

**Conclusions**

We are at a transition point in how our society thinks about consent and online privacy. That, in turn, necessitates a shift in how scientists implement citizen science. Some issues we have raised, such as credit for work and what we have called Tom Sawyer-ism, have long followed in the wake of citizen science and must be diligently guarded against. Others arise from the relative ease of online participation in citizen science efforts, as with the involvement of minor students as part of citizen science projects rather than the experiential learning of participatory science. Still others derive from borrowing social media data privacy norms with insufficient care; norms that do not rise to the standards of ethical conduct in science.  Recent incidents with companies like Cambridge Analytica and Facebook demonstrate that people feel harmed when their data is used to benefit others without their informed consent.

Profession science must acknowledge that quite possibly no citizen science project is ethically ideal. Perhaps none can be, and the best researchers can hope for is an asymptotic approach to the ideal. Projects can get closest to the ideal by not only avoiding bad practices, but also cultivating good ones.  To achieve both, teams must deliberately consider the ethical implications of consent and data privacy, and must always include the citizen scientist in ethical considerations. This does not give rise only to abstract considerations. It has concrete implications for study design, for website design, for formulating a genuinely informed consent process, and it generates the need for further theoretical and interview-based research. The recommendations we make here are being implemented with CosmoQuest.org, and should be adaptable to all projects.

This paper identifies three key ideas in viewing citizen science efforts within the context of ethical research: it is not enough to use volunteers as merely a means to an end, consent must be informed and voluntary, and people must receive credit for their efforts and understand how others (i.e. team scientists) will benefit from their efforts.

Put into practice, sites must require login, and through their registration process require agreement to a data use policy. This policy should be written in simplified English, and, if possible, utilizes graphics to tell the story visually to non-English readers. It's recognized that this will decrease the number of participants in projects as every added step causes some people to bounce away from a site without doing science, however, this is a small cost if it leads to greater long-term community trust. Additionally, sites must work to inform participants, as clearly and in-their-face as possible, what questions they are working to solve, and how scientists will benefit from their efforts. This will add additional text to webpages, and may cause sites to be deemed "less clean" or "more wordy", again, potentially causing some people to bounce away, but again, this will increase trust and understanding with the community who does stay. Finally, community members need to have an honest understanding of how they will receive credit and how their work benefits science and the scientists working with a project. There are people who would become upset to learn their efforts advance other people's degrees, promotion and tenure, or eligibility for future funding and awards. This can be mitigated through upfront honesty and by having ways that people can receive credit that they can read about before engaging. Finally, care must be taken in engaging with children and other at-risk individuals whose participation may be coerced. Every effort must be taken to prevent this, even if it reduces engagement numbers; such coercion may lead to a temporary boost in engagement numbers followed by a lifelong resentment of science by those thus forced to "volunteer" data.

No paper can do all things or be all things: there are a number of issues this paper has not been able to address here that arose during preliminary considerations and are potentially fruitful lines of conceptual and empirical inquiry. Is voluntariness undermined by the fact that particular citizen science projects, like Project Ladybug, sometimes go viral? Relatedly, do peer pressure and social status undermine voluntariness? After all, science and tech have long struggled against negative peer pressure/social esteem for science and tech. This has classically undermined voluntariness by putting young people especially in a position where they must avoid science or tech in order to be seen as feminine (for girls) and as not "too white" (for young people of color) or even to be seen as masculine and not too nerdy (for boys). But as the social status of pursuing STEM fields has shifted, and stereotypes are in flux and mix both positive and negative presumptions, citizen science faces a companion issue. Does positive social esteem or peer pressure undermine voluntariness similarly? That depends on whether it narrows the range of reasonable choices unduly, or actually opens them up. Perhaps the same mechanisms, if they lead to what philosophers call "an open future" with expanded choices (Feinberg 1980), are more acceptable than if they close off choice options. These are issues that must be pursued through not only the application of ethical theory, but also through interview-based research that allows us to understand the concerns, motivations, and lived experiences of the people attracted or recruited to citizen science projects.

Another fruitful line of inquiry is the degree to which data handling actually conforms to the promises made in informed consent materials. If the informed consent materials promise that personal data will not be shared with third parties, is this actually the case? Or is personal data shared with funders, or even sold in aggregate to recruiters or marketers in a violation of dispositional privacy?

These questions are important ones that remain to be addressed by future work. This paper provides researchers developing citizen science projects with the tools necessary to detect and to resolve some ethical issues with consent and data privacy which they might once have missed. It is enough to go on with. And go on we must: good science, after all, must also be Good science.

Footnotes

[A] Uwingo was a project that sold naming rights to craters on the Moon and Mars, and used part of the funds collected to fund science and science education projects. These names are not acknowledged by the International Astronomical Union.

# References

Aristotle. (1999). *The Nicomachean Ethics*. Trans. Terence Irwin. Indianapolis: Hackett.

Bakerman, M., et al. (2019) Motivations and Patterns of Engagement of CosmoQuest Participants. *Conference proceedings of the Astronomical Society of the Pacific*. in-press

Blumenthal-Barby, J.S. and Burroughs, H.. (2012). Seeking Better Health Care Outcomes: The ethics of using the "nudge." *American Journal of Bioethics.* 12(2): 1-10.

Bonney, R., Cooper, C. B., Dickinson, J., Kelling, S., Phillips, T., Rosenberg, K. V., & Shirk, J. (2009). Citizen science: a developing tool for expanding science knowledge and scientific literacy. *BioScience*, 59(11), 977-984.

Brandt, A. M. (1978). Racism and research: the case of the Tuskegee Syphilis Study. *Hastings center report*, 21-29.

Code, N. (1949). The Nuremberg Code. *Trials of war criminals before the Nuremberg military tribunals under Control Council Law*, 10, 181-182.

Committee on STEM Education (2018) "Charting a Course for Success: America's Strategy for STEM Education" National Science & Technology Council, Executive Office of the President of the United State

Department of Health, E. (2014). The Belmont Report. Ethical principles and guidelines for the protection of human subjects of research. *The Journal of the American College of Dentists*, 81(3), 4.

Department of Justice (2011), Code of Federal Regulations Title 28 -- Judicial Administration Retrieved from https://www.justice.gov/sites/default/files/usao-mn/legacy/2011/01/19/Remission%20Regulations.pdf

Gardner, Cory (1/6/2017). S.3084 - American Innovation and Competitiveness Act. Retrieved from https://www.congress.gov/bill/114th-congress/senate-bill/3084/text June 14, 2019Isaak, J., & Hanna, M. J. (2018). User Data Privacy: Facebook, Cambridge Analytica, and Privacy Protection. *Computer*, *51*(8), 56-59.

European Commission, 2018 reform of EU data protection rules. Retrieved from https://ec.europa.eu/commission/priorities/justice-and-fundamental-rights/data-protection/2018-reform-eu-data-protection-rules_en

Gardner, C. & Rept, S. (2017) S. 2084 American Innovation and Competitiveness Act, 114th Congress (2015-2016)

Gray, S. A., Nicosia, K., & Jordan, R. C. (2012). Lessons learned from citizen science in the classroom. a response to" the future of citizen science.". *Democracy and Education*, 20(2), 14.

Gugliucci, N., Gay, P., & Bracey, G. (2014, July). Citizen science motivations as discovered with CosmoQuest. In *Ensuring Stem Literacy: A National Conference on STEM Education and Public Outreach.* 483, 437

Faden, R. R., & Beauchamp, T. L. (1986). A history and theory of informed consent. Oxford University Press.

Feinberg, J. (1980). The Child's Right to An Open Future. In Aiken, William and LaFollette, Hugh eds. Whose Child? Children's Rights, Parental Authority, and State Power. Totowa, NJ: Rowman and Littlefield. 124-153.

Flick, C. (2016). Informed consent and the Facebook emotional manipulation study. *Research Ethics*, 12(1), 14-28.

Granville, K. (2018). Facebook and Cambridge Analytica: What you need to know as fallout widens. *The New York Times*. March 19. Retrieved from https://www.nytimes.com/2018/03/19/technology/facebook-cambridge-analytica-explained.html on June 14, 2019.

Kant, I. (2009). Groundwork of the Metaphysics of Morals. Trans. H.J. Paton. New York: Harper Perennial.

Kramer, A. D., Guillory, J. E., & Hancock, J. T. (2014). Experimental evidence of massive-scale emotional contagion through social networks. *Proceedings of the National Academy of Sciences*, *111*(24), 8788-8790.

Kerson, R. (1989). Lab for the Environment. *MIT Technology Review*. 92(1), 11–12.

Lifton, R. J. (1986). *The Nazi doctors: Medical killing and the psychology of genocide* (Vol. 64). New York: Basic Books.

Lyerly, A., Little, M., and Faden, R. 2008. The Second Wave: Toward responsible inclusion of pregnant women in research. *International Journal of Feminist Approaches to Bioethics.* 1(2): 5-22.

Masters, K., Oh, E. Y., Cox, J., Simmons, B., Lintott, C., Graham, G., Holmes, K. (2016). Science learning via participation in online citizen science. *Journal of Science Communication*, 15(3), A07.

Mill, J. S. (2016). Utilitarianism. In *Seven masterpieces of philosophy* (pp. 337-383). Routledge.

NOAA, "Cooperative Weather Observers", NOAA History. (June 8, 2006). Retrieved from https://www.history.noaa.gov/legacy/coop.html on June 14, 2019

NOAA, "History of the National Weather Service", Retrieved from: https://www.weather.gov/timeline on June 14, 2019

Pellegrino, E. D. (1997). The Nazi doctors and Nuremberg: some moral lessons revisited. *Annals of internal medicine*, 127, 307-308.

Penn, M., & Citizen, C. A. T. E. (2018, January). First results from the Citizen CATE Experiment from August 2017. In *American Astronomical Society Meeting Abstracts# 231* (Vol. 231).

Raddick, M. J., Bracey, G., Gay, P. L., Lintott, C. J., Murray, P., & Schawinski, K. (2010). Galaxy Zoo: exploring the motivations of citizen science volunteers. *Astronomy Education Review,* 9: 010103.

Roth, W. M., & Lee, S. (2004). Science education as/for participation in the community. *Science education*, 88(2), 263-291.

Shaw, D. (2016). Facebook's flawed emotion experiment: Antisocial research on social network users. *Research Ethics*, 12(1), 29-34.

Song, P. (2017). *Biomedical Odysseys: Fetal Cell Experiments from Cyberspace to China* (Vol. 12). Princeton University Press.

Thomson, J.J. (1971). "A Defense of Abortion." *Philosophy and Public Affairs* 1(1), 47–66.

Thompson, M. (2005) "Ethical Theory, 2nd Edition." London: Hodder Murray Publisher

The National Commission for the Protection of Human Subjects of Biomedical and Behavioral Research (1979), "The Belmont Report Ethical Principles and Guidelines for the Protection of Human Subjects of Research" Retrieved from https://web.archive.org/web/20040405065531/http://ohsr.od.nih.gov/guidelines/belmont.html

Utley, G. J. A. E. R. (Ed.). (1992). *The Nazi doctors and the Nuremberg Code: human rights in human experimentation: human rights in human experimentation*. Oxford University Press, USA.

Whitbeck C. (2011). Ethics in Engineering Practice and Research, Second Edition. New York: Cambridge University Press.

Yasar, S., & Baker, D. (2003). The Impact of Involvement in a Science Fair on Seventh Grade Students. Paper presented at the annual meeting of the National Association for Research in Science Teaching, Philadelphia, PA.

Young, I.M. (1990). Justice and the Politics of Difference. Princeton, NJ: Princeton University Press.

| **Citizen Science** | **Participatory Science** |
|---|---|
| Observe Eclipse Corona w Camera and submit data to Project CATE | Observe an eclipse with solar glasses and draw pictures to share at the local library |
| Report observed Meteor Shower to International Meteor Organization | Watch meteor shower with friends and point out bright meteors. |
| Report a new supernova's data to the International Astronomical Union | Photograph a new supernova and sell the image to the news |
| Document a comet's changing brightness to the Pro-Am Collaborative Astronomy Project | Note to self a comet's changing brightness at monthly star parties |

**Table 1: Citizen Science versus Participatory Science.** The key features of citizen science are the emphasis on reporting data and the potential for that data to generate new knowledge. The tasks of participatory science may be identical, such as observing meteors by eye, but in the absence of data reporting, knew knowledge for society can't be generated.

**Figure 1: Graphical Informed Consent.** We use simplified graphics and basic English to explain to volunteers how their data will be used, how they will get credit, and how others may benefit.

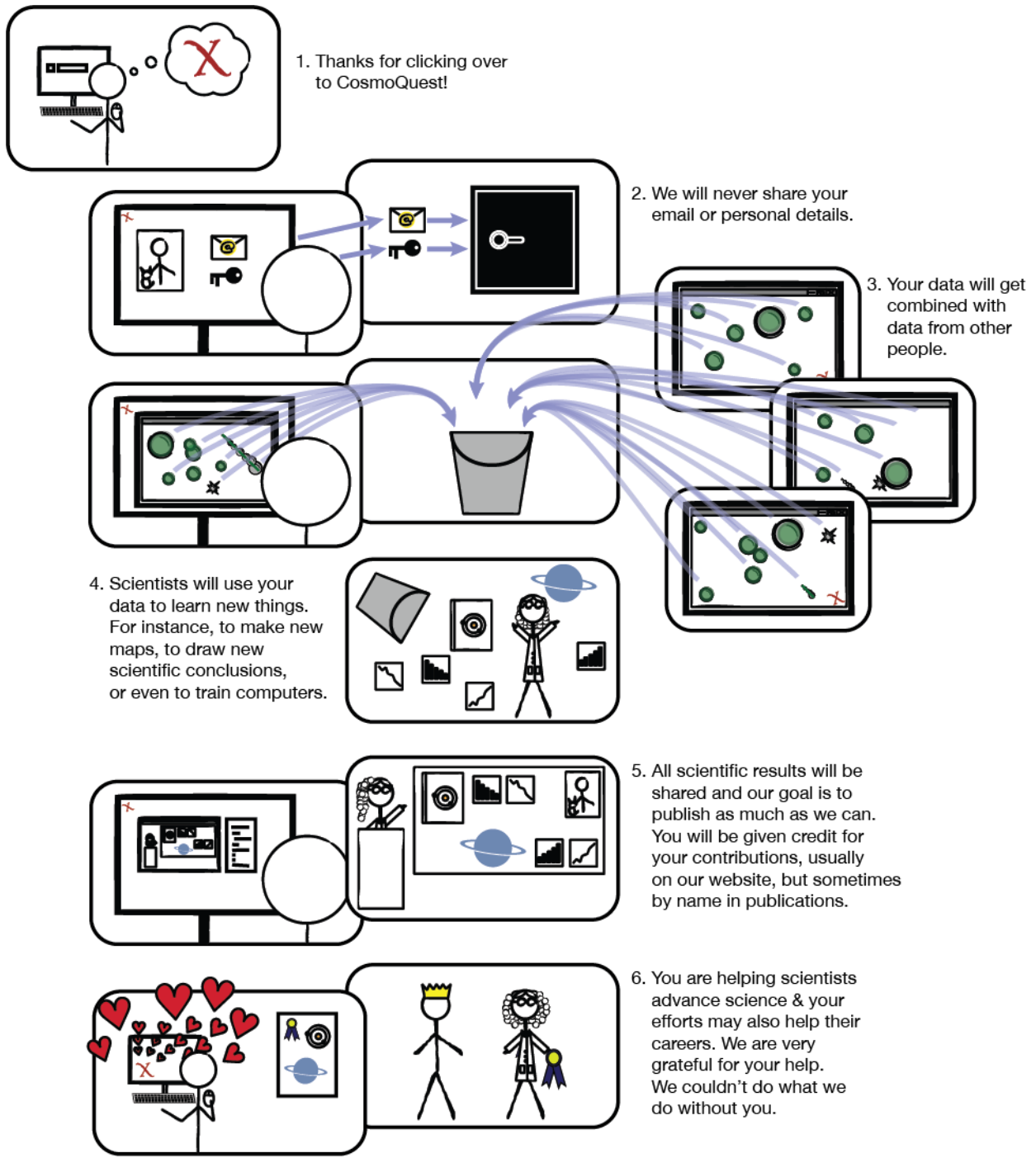